# Revisiting Strategic Cyberwar Theory – Reaching Decisive Strategic Outcome

Jan Kallberg, PhD[1]

[1]*United States Military Academy and Army Cyber Institute, Spellman Hall room 4-33, 2101 New South Post Rd, West Point NY 10996, USA*

Corresponding author: Jan Kallberg jan.kallberg@westpoint.edu

**Acknowledgement**

The views expressed herein are those of the author and do not reflect the official policy or position of the Army Cyber Institute, the United States Military Academy, the Department of the Army, or the Department of Defense.

# Revisiting Strategic Cyberwar Theory – Reaching Decisive Strategic Outcome

**Introduction**

Each strategy has a foundation – an overarching way of explaining why things are the way we see them and how to successfully reach our goals. Therefore, strategy is theory-based because theory provides an intellectual framework for predicting outcomes leading to the end goal the strategy pursues. This article will present a theory – strategic cyberwar theory – that states that the utility of strategic cyberwar is tied to the likelihood of institutional instability in the targeted nation. In an ideal scenario, the cyber attacks are systematically attacking the targeted adversary's institutions triggering the dormant entropy embedded in a nation with weak institutions. This will lead to submission to foreign policy and intent.

The current alternative to strategic cyberwar theory is to unsystematically attack the adversary with cyber attacks where exploitation opportunities occur, which is likely to degrade parts of the information infrastructure, but it will not reach any strategic goals. If an adversarial society is unaffected by a cyber conflict, the conflict itself has not reached a decisive outcome, and results only in tit-for-tat game or stalemate. *How do we achieve a decisive outcome? How do we win a cyberwar?*

In strategic cyberwar theory[1], the concept is to cyber attack the core of the institutional framework of the adversarial nation in pursuit of destabilization. The theory's predictive power is strongest when applied to targeting authoritarian regimes, and dysfunctional failed democracies, because the common tenet is weak institutions.[2] Fully functional democracies or constitutional monarchies, on the other hand, have in cyberwar a definite advantage because these advanced societies have stabile and by their citizenry accepted institutions. Nations openly adversarial are in most cases totalitarian states that are also close to entropy and disorder. The reason why these totalitarian states are under their current regime is suppression of the popular will. Any removal of the pillars of suppression, by destabilizing the regime design and institutions that make it functional, will release the popular will. A destabilized and possibly imploding Iranian regime is a far more tangible threat to the ruling Iranian elite than military information subsystems being hacked.

Theory is an overarching way of combining ideas, phenomena, and facts, in a generalized form, to seek to explain specific outcomes. Theory's strongest tenet is predictability. Theory can serve as a guidance to prepare for future events and ensure that these outcomes are favorable.

In a militarized Internet it is convenient to lean towards traditional military theory transposed into cyber. Traditional military theory applied in cyber struggles with four challenges – anonymity, object permanence[3], measurable results, and rapid digital execution. In a Clausewitzian world, these challenges were non-existent. First, the enemy was clearly marked. A state of war was declared. A French Napoleonic general overlooking the battle could clearly distinguish a thin red line of British troops waiting from the advancing French Guards in blue uniforms. There was a basic understanding of who were the parties in the conflict, their past actions, and the strategy that drove their action. Next challenge for traditional military strategy is object permanence.[4] The general could march with his armies to a point where he the next day will fight the battle and on a map lay out his course of action. The landscape would be intact the next day, the roads had not moved, and the hills stood where they should. If there is no object permanence, maneuvering concepts[5] becomes irrelevant because the maneuver is a positioning to increase the opportunity for success – and if we are unable to relate in time and space maneuvering is nullified. A third challenge is quantifiable results. Measurable results are needed as information for further decision-making and battle assessment. Cyber lacks the feedback loop of quantifiable results and with no measure of effectiveness the next move in traditional military theory cannot be executed according to the theory – because it relies on a chain of events leading to a decisive moment. The famous OODA-loop (Observe, Orient, Decide, Act) for decision making does not work when several steps are missing.   Computers at war does not engage at human speed, the future engagements occur at computational speed would eradicate any influence of real-time human leadership. The uniqueness of cyber removes the predictive power of traditional military strategy.

**Going from the Unknown to the Known**

If battle results cannot be quantified, there is no object permanence, and the assumed enemy is anonymous, and the battle occur at computational speed – any grander battle strategy is becoming inferences about the unknown. Strategic cyberwar theory[6] utilize the thinking of Bertrand Russell in his version of Occam's razor; "Whenever possible, substitute constructions out of known entities for inferences to unknown entities."[7] Occam's razor is named after the medieval philosopher and friar William of Ockham who stated that in uncertainty the fewer assumption the better and pursuing simplicity by relying on the known until simplicity could be traded for greater explanatory power. What is the basic knowledge we have with least uncertainties?

Societies are engaged in conflicts. The main building block for any society is institutions. The institutional resilience varies by nation, from stabile democracies to totalitarian states on the brink to entropy, and the destabilization effort need to reach the whole society with an intensity reaching beyond the targeted nation's resiliency.

If institutions fail, society will be destabilized and weakened. A destabilized society collapses or is subdued to foreign power. Following the stated known, strategic cyberwar theory seeks to explain how an adversarial society can be destabilized and subdued by a major cyber campaign. Cyberwar must be quickly executed and unprecedented in the aim of the attack. The reason is the opportunity to shock the targeted society and in the same moment avoid adaptive behavior that mitigates the damages from the attacks. The rapid execution denies the targeted nation the opportunity to create defensive measures and evaporate any possibility to strategically lead the cyber defense.

An attack will fail to destabilize the targeted society if the institutions are intact after the attack – or able to operate in a degraded environment. Therefore, it is important to ensure that the attack is of the magnitude that is forces the targeted society over the threshold to entropy.[8]

The risk of seeing the world emerging as a mechanical part of the world before assumes that the environment it is submerged in will not change. The trap that is created by path dependency and tradition can be presented by another word – assumption.
The main risk, I argue in the current cyber discourse focusing on cyber as purely an enabler of joint operations is the numerous assumptions built in and these assumptions are products of tradition burdened perceptions, lacking understanding of the reversed asymmetry of the conflict, where a state can attack a domestic public entity and down to the single citizen, the absence of object permanence, absence of acceptance of the rapid time frame interchanges will occur, and the impact of artificial intelligence in combination with automated harvest of vulnerabilities.

If cyber is limited to only be enabler, other operational intent will drive the execution towards the strategic goal. I argue that cyber offers a strategic opportunity that will grow the coming decades.

**Strategic Cyberwar Theory**

If states seek to conduct decisive cyberwar it will not be achieved by anecdotal exploits, but instead by launching a systematic destabilizing attacks on the targeted society. In strategic cyberwar theory the intellectual works of Dwight Waldo are utilized, meanwhile Waldo studied the theoretical underpinnings to maintain government institutional sustainability and stability, strategic cyberwar theory turn these theories upside down to create entropy and destabilization. The systematic approach seeks to use institutional weaknesses, popular sentiment, and underlying opposition to the targeted government as force multipliers to the effect. The targeting can induce a sense of lack of control and failure to safe-guard their citizenry. [9] A nation, or any societal structure, is organized through institutional arrangement. These arrangements require a set of basic functionalities to operate within the institution to ensure the continued stability and functionality. Institutions make a state stable, a government sustainable and functional, even in a degraded environment. Each country is unique in its institutional arrangements and the societal importance of these arrangements.

A systematic institutional cyber attack can be visualized as the collapse of a building built with prefabricated elements, such as a parking garage, on a framework of concrete beams, pillars and decking. If pressure is distributed evenly over the construction, as an early morning rain, there is no risk for a collapse. The building is safe. If instead the energy is concentrated on one or a set of the bearing elements of the building it will collapse. Dwight Waldo's theoretical work explained what makes a nation state stable.[10] The strategic cyberwar theory turns Waldo's accepted theories upside down, so instead of upholding the functionality of the targeted society it seeks to swiftly destabilize the targeted society. Waldo named five factors – legitimacy, authority, knowledge management, bureaucratic control, and confidence. Authority could then be external authority, by leading or in some cases suppressing a people, and internal authority within the bureaucracy and political structure.

**Waldo's Five Pillars for Societal Stability**

Waldo's five factors legitimacy, authority, knowledge, control, and confidence summarize the pillars of any society and government. If a major automated attack can undermine these pillars the targeted society are either weakened or at risk to implode. Legitimacy includes not only that the government is legally legitimized, but capable and focused on an intention to deliver the "good society" or in a dictatorship "acceptable society". Legitimacy is a sliding grey-scale and cannot be seen as a value that the society either has or not. Authority is the ability to implement policy. In a democracy or constitutional monarchy, it requires the acceptance of the people based on rationalism, expectations of public good, ethics, and institutional contexts. Knowledge is institutional knowledge, the ability to arrange and utilize knowledge within the bureaucracy since coordination is the major challenge in knowledge management. Control is the ability to control what we want to control in the bureaucracy. Confidence is the trust people have that government delivers the expected benefits and the removal of fear for the future. According to Waldo, feelings of vulnerability and fear of future events are the absence of confidence in government.

These five factors are the framework that holds a government together. If depleted and removed the absence of the factors will disintegrate government. In strategic cyber warfare it is pivotal to remove any of these pillars, leading to the collapse of the other, and damage the targeted society.

Legitimacy concerns not who can lead but who can govern. Waldo believed that we need faith in government; for government to have a strong legitimacy it has to project, deliver, and promise that life would be better for citizens. For a major automated attack seeking to damage legitimacy in a state it has to darken the future for the population, create a notion that the leadership are unable to govern the country to a better position for the individual, and that the current regime create undue burdens for the citizens.

Authority in totalitarian regimes can be summarized as acceptance for the moment. Authority and hierarchy are linked when the structure of the hierarchy determines the authority of a specific position. If there is no hierarchy, there is no leadership that can be held accountable for its actions; with no accountability, any organization would fall into entropy and anarchy.

One of the major challenges for modern government is knowledge management. If public administrators are unable to organize knowledge and information, the public is left with the impression that the government is incompetent. This is an indirect challenge to authority and could lead to societal entropy. A modern society generates massive amounts of information at all levels. If a lack of knowledge and coordination affects citizens, it undermines their perception of how well government is working. Cyberattacks on knowledge management will cripple the bureaucracy and anger the population.

Complex organizations have problems with bureaucracy as they grow in size and complexity. Control can also be lost because of a lack of coordination among agencies, local and state governments, and other stakeholders. When government does not have control across organizations, jurisdiction is lost. As bureaucracy expands, so do the control issues since control requires coordination. Control issues also arise through unintentional errors. If control is lost corruption, favoritism, public theft, and popular discontent will follow.

Waldo connected the words secure and confidence when he described his confidence problem. When people feel secure, they have confidence and are optimistic about the future; they trust government will provide support. Confidence for Dwight Waldo was trust in government to deliver the society it promised. Confidence means that the future is perceived to be brighter than the past; legitimacy and authority is defined in the present, confidence is forward looking. Emerging events of scarcity and competition for public resources is harmful to confidence in government, because it challenges future ability to serve citizens. Signs of systematic failure and projected inability will harm the citizenry's ability to maintain confidence in government. If several of these five pillars for societal stability are damaged it can lead to system collapse and the downfall of a regime.

**Examples of Targeting**

Strategic cyberwar theory predicts the weaknesses of the targeted government – in the pursuit of remotely initiated regime shift or submission to foreign power. These weaknesses are identified in each society based on the societal characteristics and tenets. Once the weaknesses are identified they are aligned with the theory and operationalized to targeting. The attack in these sectors is likely unexpected by the targeted nation. These targets selected by strategic cyberwar theory differ in several cases from the traditionally prioritized assets for national cyber security and information assurance, such as military, defense-industrial, diplomatic, and executive information assets.

An example is a one-party dictatorship that has successfully politically survived by providing consumption and financial progress to the vocal part of its citizenry. The one-party dictatorship has a set of unique tenets. The government is highly centralized and rule from a set of iconic building structures. Building sector and real estate has been booming where money is funneled through informal banking institutions, which operate outside of the party-controlled system, and the money have provided mortgages. The informal banking sector is an inviting target opportunity.[11] All banks are a database that sorts out who owes who and who can establish a demand. A database can be destroyed or corrupted. The unleashed entropy by systematically bold and swiftly attacking the informal banking system is likely to be extremely high.

The one-party dictatorship relies on pay-outs to loyalists, which then becomes a target with corrupted payments. The identification process is vital for strategic cyberwar theory.

| EXEMPLE OF TARGETING MATRIX - ADVERSARIAL ONE PARTY DICTATORSHIP ||
|---|---|
| Waldo's Five Factors | Example of Targets |
| Legitimacy | Deny Electricity for Iconic Administrative Centers |
| Authority | National Police Information Sharing Dissemination of Loyalist Informers' Personal Data |
| Institutional knowledge | Real-Estate/Cadastral Data Corrupting Land Ownership Information Destruction of Permit Databases |

| Control | Corruption of Government Salary Pay-Outs |
| | Degrade the Blocking Operations that |
| | Prevent Access to the Complete Internet |
| Confidence | Informal Banking Institutions |

For the attacker, the keys to success for implementation and use of strategic cyberwar theory are the pre-planning and mapping of the institutional design and weaknesses of the future targeted society. The swiftness is important to avoid any adaptive behavior in the targeted state. The modernized countries have a corporate and federal culture of rapid patch management, following the different information security management structures and protocols in place, but the potential adversarial nations are less able to patch their networks in time. The rapid execution of the attack ensures that the feedback loop generated by the attack do not generate a healing of the systems – due to the fact that the patch management in place is too unstructured, driven by commands instead of delegated initiative, and therefore lack rapid response.

**Conclusions**

The core claim is that cyber will be a way to reach geopolitical goals in the future by destabilizing adversarial nations. Strategic cyberwar theory is a tool to exploit the weaknesses in adversarial states – if needed. Eventually, cyber will be able to catapult adversarial countries into entropy by creating a system chock to the institutional framework that holds these countries together. As stated, traditional military theory applied in cyber struggles with four challenges – anonymity, object permanence, measurable results, and rapid execution. If traditional military thinking is utilized to formulate a strategy, it is likely that the result would aggregate spurious assumptions and remove the opportunity for decisive offensive cyber operations as a geopolitical toolset. Strategic cyberwar theory is viewing the adversarial nation as a framework of institutional arrangements instead of a set of military assets and digital networks. The institutional frameworks are likely to be less well defended as the industrial-military complex, but when destabilized these frameworks removes the underpinnings for the adversarial regime leading the way to a decisive end to the cyber conflict. The theory also argues that the attacks have to occur in a limited time frame to ensure system chock in the targeted society.

---

[1] Jan Kallberg. "Strategic cyberwar theory-A foundation for designing decisive strategic cyber operations." *Cyber Defense Review* 1, no. 1 (2016): 113-128.

[2] Paul Brooker, *Non-Democratic Regimes: Theory, Government and Politics* (New York: Palgrave Macmillan 1994).

[3] Jan Kallberg and Bhavani Thuraisingham, 'Cyber Operations: Bridging from Concept to Cyber Superiority', *Joint Forces Quarterly* 68 (2013).

[4] Jan Kallberg, and Thomas S. Cook. "The unfitness of traditional military thinking in cyber." IEEE Access 5 (2017): 8126-8130.

[5] Applegate, Scott D, 'The Principle of Maneuver in Cyber Operations', In Cyber Conflict (CYCON), 2012 4th International Conference on, pp. 1-13. IEEE, 2012.

[6] Jan Kallberg, Bhavani Thuraisingham, and Erik Lakomaa, 2013, 'Societal Cyberwar Theory Applied the Disruptive Power of State Actor Aggression for Public Sector Information Security', Presented at and published in *Proceedings from the 2013 IEEE European Intelligence and Security Informatics Conference (EISIC 2013)*.

[7] John Shand, *Philosophy and Philosophers: An Introduction to Western Philosophy*, (Montreal: McGill-Queen's Press-MQUP 2002).

[8] Jan Kallberg, and Adam B. Lowther, 'The return of Dr. Strangelove', *The Diplomat*, Aug. 20, 2012.

[9] Jan Kallberg, and Rosemary A. Burk, 'Failed cyberdefense: The Environmental Consequences of Hostile Acts, *Military Review*. (May-Jun. 2014), 22-25.

[10] Dwight Waldo, *The Enterprise of Public Administration* (Novato: Chandler & Sharp 1980).

[11] The Economist, 'Shadow banking in China - The Wenzhou experiment', Apr. 7, 2012, http://www.economist.com/node/21552228>